\documentclass[runningheads,hidelinks]{llncs}

\usepackage[T1]{fontenc}
\usepackage{graphicx}
\usepackage{booktabs}
\usepackage{url}
\usepackage{amsmath}
\usepackage{orcidlink}
\usepackage{microtype}

\begin{document}

\title{Threat Amplified, Blame Restrained: LLM-Assisted Media
Framing Analysis of the 2026 Bangladesh Measles Outbreak}

\titlerunning{Media Framing of the 2026 Bangladesh Measles Outbreak}

\author{Shahan Ahmed}

\authorrunning{S. Ahmed}

\institute{Independent Researcher\\
\email{shahan24h@gmail.com}}

\maketitle

\begin{abstract}
How news media frame and emotionally code a public health emergency shapes public risk perception and trust, yet outbreak-coverage dynamics remain understudied for low- and middle-income countries (LMICs). We examine sentiment and stance in English-language Banglad-eshi coverage of the 2026 measles outbreak---the country's most severe in two decades, with over 97{,}000 suspected cases and 600 deaths across 61 of 64 districts, unfolding after the 2024
change of government and a 2024--2025 vaccine stockout. Using the Internet Archive, we build a reproducible corpus of 403 headlines from seven national outlets (396 in-window in 2026), label them for binary sentiment and four-way stance via a large language model under a locked codebook, and validate against a two-coder
human-adjudicated gold standard ($n{=}153$; Cohen's
$\kappa{=}0.89$ stance, $0.75$ sentiment). Aligned to the DGHS epidemic curve, coverage grew significantly more negative (56\%$\to$88\% negative; Cochran--Armitage $z{=}4.12$, $p{<}.001$) and risk-amplification framing 
intensified (44\%$\to$84\%; $z{=}4.15$, $p{<}.001$). Media negativity lagged incidence, tracking cumulative mortality. Contrary to the political backdrop, blame remained a minority frame ($\sim$9\% overall) and was
overwhelmingly systemic (32 of 37, 86\%) rather than directed at named actors. The pipeline offers a scalable, transparent method for LMIC outbreak-media analysis; Bangladeshi coverage amplified threat far more than it assigned political blame.

\end{abstract}

\section{Introduction}

Media framing of a health emergency shapes what publics fear, whom
they blame, and whether they act. Yet most computational studies of
outbreak coverage focus on high-income settings and dominant
platforms, leaving the framing dynamics of low- and middle-income
countries---where epidemics carry the highest burden---
underexamined. The 2026 measles outbreak in Bangladesh offers a
consequential case: the country's most severe measles crisis in two
decades, unfolding under a politically contested interim
administration and a documented 2024--2025 vaccine stockout,
producing more than 97{,}000 suspected cases and 600 deaths across
61 of 64 districts~\cite{dghs2026}. Whether coverage of such a
crisis foregrounds systemic critique, personal political blame, or
technocratic reassurance is both a substantive and a
methodological question.

The outbreak also carries an important political dimension. From
August 2024 to February 2026, Bangladesh was governed by an interim
administration under Chief Adviser Muhammad Yunus, installed
following mass protests that led to the resignation of Prime
Minister Sheikh Hasina~\cite{idea2024bangladesh}. This period was
marked by political turmoil that disrupted routine health
programming; the quadrennial measles--rubella mass vaccination
campaign due in 2024 did not take place, and vaccine and medicine
procurement was slowed by administrative and funding
delays~\cite{bmj2026bangladesh,prothomalo2026shortage}. The 2026
outbreak thus unfolded against a backdrop in which the immediate
causes---programme disruption, stockouts, workforce vacancies---
were politically legible, raising the empirical question of whether
media framing tracked those political dynamics or the
epidemiological crisis itself.

This paper contributes both a substantive analysis and a
transferable method. Substantively, we ask how sentiment and
framing evolved across the outbreak's phases, and whether coverage
tracked epidemiological reality or its cumulative human cost.
Methodologically, we demonstrate a validated LLM-annotation
pipeline anchored to permanent Internet Archive URLs, addressing a
persistent reproducibility gap in news-corpus research. We report
strong LLM--human agreement on stance ($\kappa{=}0.89$),
statistically significant intensifications of negativity and threat
framing, and a striking negative finding: despite the political
context, blame remained sparse and overwhelmingly systemic.

\section{Related Work}

\subsection{Issue-Attention and Framing of Health Crises}

Downs's issue-attention cycle~\cite{downs1972} describes how public
problems move from a pre-problem stage through alarmed discovery,
recognition of costs, gradual decline of interest, and post-problem
phases. Applications to disease coverage document analogous
trajectories in outbreak reporting, with volume spiking on onset
and coverage progressively foregrounding severity as consequences
accumulate~\cite{shih2008,ogbodo2020}. A related literature on
framing distinguishes threat-focused (risk-amplifying) from
reassurance-focused frames, and systemic from individualizing blame
attributions---distinctions that carry policy weight because they
shape which actors are held accountable~\cite{schafer2019}.

Framing scholarship also distinguishes attributions of
responsibility along a systemic--individual axis: whether coverage
locates fault in structural failures (policy, programme design,
supply chains) or in identifiable actors (officials,
agencies)~\cite{iyengar1991}. In politically charged health crises,
this distinction shapes downstream accountability---who is asked to
answer for the crisis, and what kinds of remedies are demanded.

\subsection{LLMs as Annotators for Computational Social Science}

Recent work has established that large language models can produce
annotations comparable to trained human coders on many text-
classification tasks, at a fraction of the cost~\cite{gilardi2023,
heseltine2024}. Stance detection is generally harder than
sentiment~\cite{walker2025}: it requires distinguishing what a
text \emph{claims about} its subject from mere valence, and
performance varies more across models and prompts. Best practice
therefore treats LLM labels as candidate annotations that must be
validated against a human gold standard using standard inter-
annotator statistics before being used at scale.

\subsection{The 2026 Bangladesh Measles Outbreak}

The 2026 outbreak followed years of declining measles-rubella
vaccination coverage and a 2024--2025 supply disruption, with a
declared onset on 15 March 2026 and rapid geographic spread
requiring an emergency nationwide vaccination campaign launched in
early April~\cite{dghs2026,who2026}. Analytical accounts have
emphasized structural drivers---programme
disruption, delayed diagnostics, and immunization gaps---as
proximate causes~\cite{hossain2026}, while public commentary has
sometimes framed the crisis politically. Whether English-language
media reflected the structural or the political framing is an open
empirical question this paper addresses.

The 2026 Bangladesh outbreak sits within a global pattern of
measles resurgence: measles vaccination coverage and herd-immunity
levels worsened across most WHO regions between 2019 and
2023~\cite{plansrubio2025}. Recent domestic analyses have
emphasized programmatic drivers---the missed 2024 mass campaign,
workforce vacancies, and disrupted procurement---as proximate
causes of the Bangladeshi resurgence~\cite{hossain2026,who2026don598},
framing the crisis as a public-health-system failure rather than
an isolated epidemiological event.

\section{Methods}

\subsection{Sampling Frame and Search Terms}

The frame comprises seven national English-language Bangladeshi
newspapers---\emph{The Daily Star}, \emph{Dhaka Tribune},
\emph{New Age}, \emph{bdnews24}, \emph{The Business Standard},
\emph{The Financial Express}, and \emph{Prothom Alo}
(English)---selected \emph{a priori} for national reach, original
English reporting, and archival coverage of 2023--2026. Three
candidate outlets (\emph{The Daily Observer}, \emph{banglanews24
English}, \emph{Bangladesh Today}) were excluded on documented
technical grounds: opaque numeric-identifier URLs or insufficient
2023--2026 archival coverage. The search vocabulary was restricted
to \emph{measles} (case-insensitive, word-boundary); rubella-only,
MMR-only, and generic vaccination items were excluded. The window
spans 2023-01-01 to 2026-12-31; pre-2026 coverage was retained
without selection on outcome to establish a baseline against which
2026 dynamics could be interpreted.

\subsection{Archival Retrieval and Extraction}

We queried the Internet Archive's CDX Server API for every archived
capture of each outlet domain within the window, consolidating
$\sim$2.58 million unique URLs. Candidate articles were identified
by regex-matching the search term against URL slugs, exploiting
these outlets' practice of embedding headline text in article
paths. Prothom Alo, whose URLs use opaque random identifiers,
required an adapted procedure preserving the same scope: we
fetched all Prothom Alo \texttt{/story/} snapshots captured in the
2026 window (4{,}020 pages after \texttt{/amp/} deduplication) and
applied the same regex to \emph{extracted headline text} rather
than the URL slug. For each candidate, headlines and publication
dates were extracted from a cascade of sources (Open Graph tags,
HTML \texttt{h1}, JSON-LD, \texttt{<time>} elements), with the
Wayback capture date used as a documented fallback. Six
bad-extraction stubs and one non-Bangladesh headline were removed.
The final corpus comprises 403 headlines across seven outlets, 396
in-window in 2026.

\subsection{Codebook and Categories}

Two axes were coded per headline. \emph{Sentiment}: negative,
neutral, or positive; later collapsed to binary (negative
vs.\ non-negative) after human coders diverged on the
neutral/positive boundary. \emph{Stance}: \emph{risk\_amplification}
(severity/scale/spread/
threat, including all quantified death and
case counts), \emph{blame\_attribution} (fault assigned to a named
actor---\emph{individual}---or to a systemic failure such as a
stockout or lapsed campaign---\emph{systemic}),
\emph{reassurance} (response, control, campaigns, recovery), and
\emph{neutral\_informational} (procedural announcements,
explainers, meta-commentary). Two locked decision rules resolve
boundary cases: (i) quantified harm is coded as risk-amplification
even absent overt threat words; (ii) blame is strict---mere mention
of a shortage, probe, or crisis without asserting fault is
\emph{not} blame. Codebook v1.1 was calibrated on 20 headlines and
locked before full labeling.

\subsection{LLM Annotation}

Each headline was labeled by Anthropic Claude Haiku 4.5
(\texttt{claude-haiku-4-5-
20251001}) with $T{=}0$, a fixed system
prompt from the codebook, and five few-shot examples covering all
stance classes. Responses were JSON-parsed and validated;
malformed outputs were retried with exponential backoff. Of 324
initial-corpus items, 321 (99.1\%) were labeled first-pass and 3
on retry; the 79 Prothom Alo items were labeled subsequently under
the identical model, prompt, and codebook, yielding a
consistently-annotated 403-item corpus.

\subsection{Human Validation}

A stratified sample of 160 items was drawn from the initial
corpus, oversampling blame to ensure adequate representation.
Two independent coders labeled the sample without seeing each
other's labels or the LLM's; six bad-extraction stubs and one
US headline were removed, leaving 153 items. Three-way sentiment
showed poor inter-annotator agreement ($\kappa{=}0.37$), driven
by disagreement on the neutral/positive boundary for
response-oriented headlines; collapsing to binary yielded
$\kappa{=}0.78$. Stance agreement was moderate ($\kappa{=}0.48$),
concentrated in the reassurance/neutral-informational and strict-
blame boundaries. All 67 stance disagreements were adjudicated
by a third analyst applying the codebook's decision rules
consistently, producing a rule-based consensus gold standard.
Against this gold standard the LLM achieved $\kappa{=}0.89$ on
stance (accuracy $0.93$) and $\kappa{=}0.75$ on binary sentiment
(accuracy $0.89$). The LLM slightly under-detected blame
($\sim$17\% conservative), so blame counts below are treated as
lower bounds.

\subsection{Temporal Analysis}

Full-corpus LLM labels were aggregated by ISO week and by four
outbreak phases: emergence (Mar), escalation (Apr), peak (May),
decline (Jun+). Weekly counts and stance shares were aligned
against the DGHS epidemic curve. Three inferential tests were
performed: (i) chi-square on the stance$\times$phase 4$\times$4
table with Cram\'er's V; (ii) Cochran--Armitage trend on the
share of negative-sentiment headlines across ordered phases;
(iii) the same trend test on risk-amplification share. Because
blame-by-phase cells were small (4/12/20/1), blame is reported
qualitatively.

\section{Results}

\subsection{Corpus and Coverage Dynamics}

\begin{figure}[t]
\centering
\includegraphics[width=0.95\linewidth]{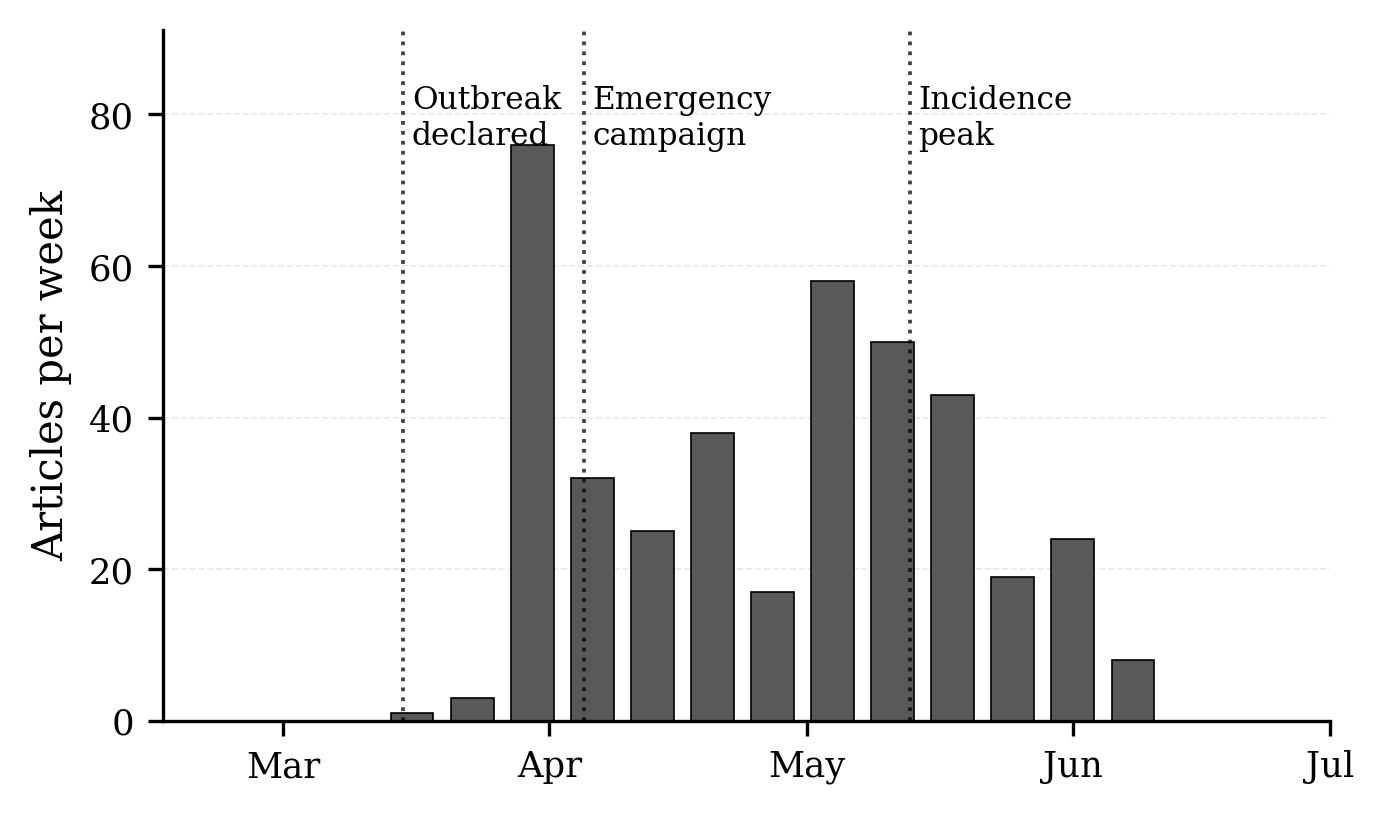}
\caption{Weekly measles article volume across the 2026 outbreak
window, with three dated reference points: outbreak declared
(15 March), emergency measles--rubella vaccination campaign
launched ($\sim$5 April), and DGHS-reported incidence peak
(week of 7--13 May). The emergence--escalation--peak--decline
trajectory tracks Downs's issue-attention cycle.}
\label{fig:volume}
\end{figure}

Measles was effectively absent from English-language coverage
before 2026---only 7 headlines predate that year, consistent with
Downs's pre-problem stage. Article volume then surged: 32
headlines in March, 160 in April, 172 in May, and 32 in a
partially archived June (Fig.~\ref{fig:volume}). Because Wayback
captures of recent pages accrue over time, the June decline
reflects both falling attention and incomplete archiving.

\subsection{Sentiment Became Monotonically More Negative}

\begin{figure}[t]
\centering
\includegraphics[width=0.95\linewidth]{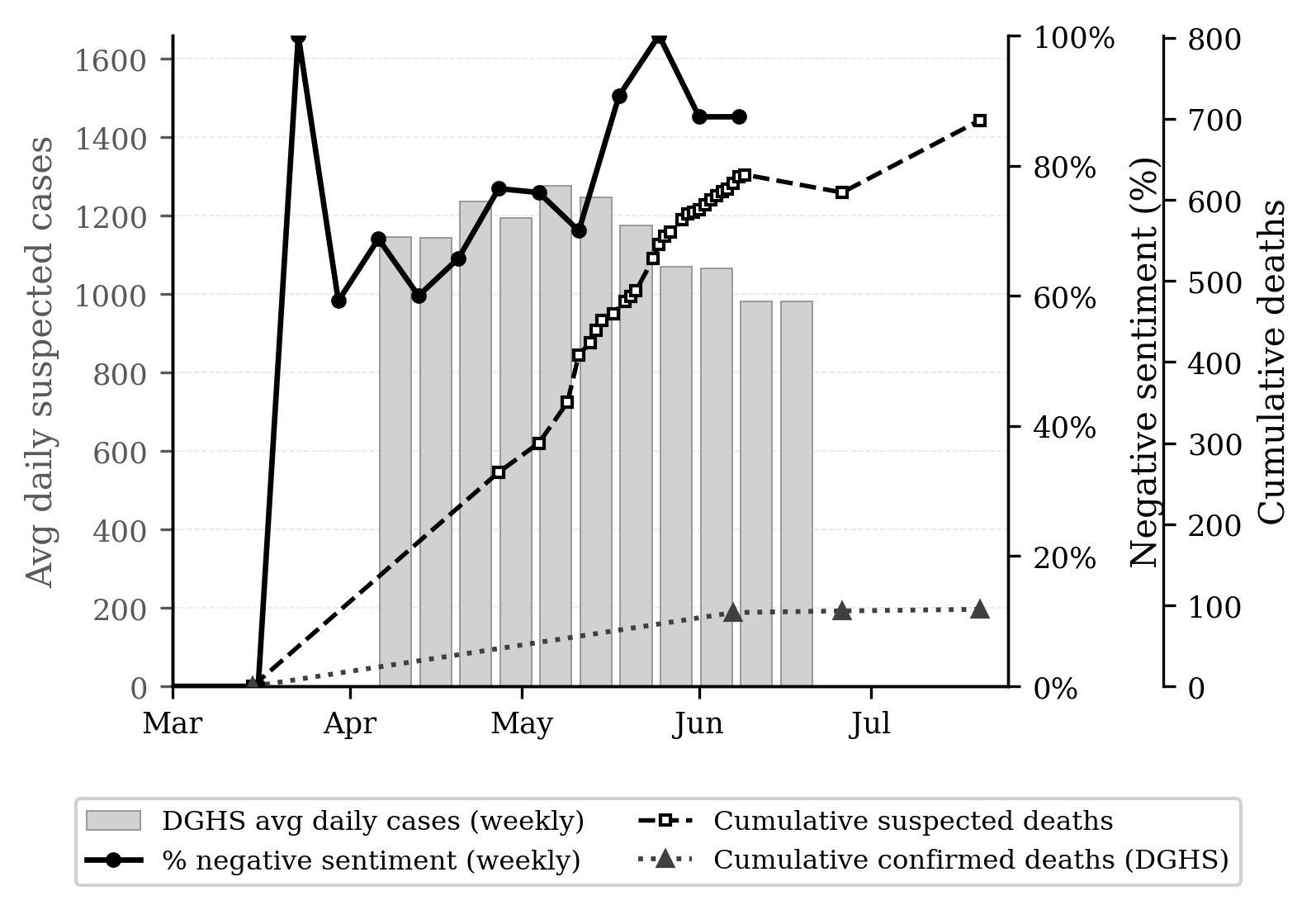}
\caption{Media negativity tracks cumulative mortality, not incidence.
Weekly percentage of negative-sentiment headlines (solid line, inner
right axis) rises monotonically. Cumulative suspected deaths (dashed
line, outer right axis) combine headline-reported daily figures with
DGHS bulletin anchors; cumulative confirmed deaths (dotted line)
reflect laboratory-verified subsets from DGHS bulletins. The DGHS
weekly average-daily suspected-case curve (grey bars, left axis)
peaks during 7--13 May and recedes through June and July, even as
media negativity and cumulative deaths continue to rise.}
\label{fig:lag}
\end{figure}

The share of negative headlines rose from 56\% (March) to
65\% (April), 80\% (May), and 88\% (June); Cochran--Armitage
$z{=}4.12$, $p{<}0.001$. Notably, negativity continued to climb
into June even as the epidemic curve receded
(Fig.~\ref{fig:lag}): DGHS reported average daily cases peaking
at 1{,}277 during 7--13 May and declining to 981 by
mid-June~\cite{dghs2026}. Media negativity thus lagged incidence,
tracking the still-rising cumulative death toll (which passed
600 by early June and reached 698 suspected deaths by late
July). Emotional framing appears anchored to cumulative human
cost rather than to real-time epidemiological improvement.

\subsection{Framing Shifted Toward Risk Amplification}

Stance composition differed significantly across phases
($\chi^2(9){=}34.44$, $p{<}0.001$; Cram\'er's V${=}0.17$).
Risk-amplification framing intensified from 44\% (March) to
56\% (April), 69\% (May), and 84\% (June) (Cochran--Armitage
$z{=}4.15$, $p{<}0.001$; Table~\ref{tab:phases}). A direct
emergence-vs-peak contrast confirmed a highly significant shift
($\chi^2(3){=}16.60$, $p{<}0.001$).

\begin{table}[t]
\centering
\caption{Stance composition (\%) and negativity by outbreak phase.
Cell counts are given as $n$ headlines per phase.}
\label{tab:phases}
\begin{tabular}{lrrrrrr}
\toprule
Phase & $n$ & \% neg & \% risk & \% blame & \% reass. & \% neut. \\
\midrule
Emergence (Mar) & 32  & 56 & 44 & 12 & 9  & 34 \\
Escalation (Apr)& 160 & 65 & 56 & 8  & 22 & 15 \\
Peak (May)      & 172 & 80 & 69 & 12 & 11 & 9  \\
Decline (Jun+)  & 32  & 88 & 84 & 3  & 3  & 9  \\
\bottomrule
\end{tabular}
\end{table}

Reassurance followed a rise-and-collapse pattern: scarce during
emergence (9\%), it spiked to 22\% in April coinciding with the
nationwide vaccination campaign launch ($\sim$5 April), then fell
back to 11\% (May) and 3\% (June). The ``managed-response''
narrative did not persist as mortality climbed.

\subsection{Blame Was Sparse and Predominantly Systemic}

Contrary to expectations from the political context, blame was a
consistently minor frame: 12\%, 8\%, 12\%, and 3\% across phases
(raw counts 4, 12, 20, 1). Because several phase$\times$stance
cells contained expected counts below 5, and because the LLM
under-detects blame, we report blame's temporal trajectory
qualitatively rather than testing for significance.

Two features are nonetheless robust. First, blame was never the
dominant frame in any phase; risk-amplification exceeded it by a
wide margin throughout. Second, where blame appeared, it was
overwhelmingly systemic: 32 of 37 blame headlines (86\%) attributed
fault to policy or programme failures (stockouts, lapsed campaigns,
diagnostic gaps), while 5 (14\%) named individual political
actors. Bangladeshi English-language coverage thus framed the
crisis chiefly as an intensifying health threat attributable to
structural failures, not as an instrument of partisan attribution.

\section{Discussion}

\subsection{Interpreting the Trajectory}

The trajectory maps onto Downs's issue-attention
cycle~\cite{downs1972}: near-absence pre-outbreak (pre-problem),
volume surge on onset (alarmed discovery), and intensifying
risk-amplification as consequences accumulate (recognition of
cost). The reassurance peak coincident with the vaccination
campaign reflects the appearance of institutional response, but
its failure to persist---displaced by threat framing as mortality
climbed---suggests that in a high-mortality outbreak,
managed-response narratives are fragile and subordinate to
accumulating human cost.

The lag between media sentiment and the incidence curve is
theoretically notable. Whereas topic-based studies typically find
attention tracking case counts, our affective measures tracked
cumulative mortality: negativity continued rising into June as new
cases fell. This suggests that emotional framing anchors to
visible cumulative cost rather than to real-time epidemiological
signals---with implications for risk communication, since public
risk perception shaped by such coverage may remain elevated after
the epidemiological threat begins to subside.

\subsection{Threat Amplification Over Political Blame}

Perhaps the most consequential finding is what did \emph{not}
occur. Given a change of government and a documented vaccine
stockout, coverage might plausibly have become a site of political
blame. Instead, blame remained a minority frame in every phase,
and where it appeared it was overwhelmingly systemic. This pattern
contributes to the systemic-versus-individualizing frame literature
by showing that even a politically charged LMIC health crisis can
be covered predominantly through a systemic-failure lens, and it
cautions against assuming politically salient outbreaks
necessarily produce politically personalized coverage.

\subsection{Interpreting the Restraint on Political Blame}

Given the political backdrop---an interim administration whose
tenure coincided with the missed 2024 campaign, delayed
procurement, and reported health-workforce vacancies of up to 45\%
in more than half of districts by 2025~\cite{prothomalo2026shortage}
---the sparseness of individual, actor-directed blame in our corpus
(5 of 37 blame headlines, 14\%) is notable and warrants
interpretation. Several non-exclusive explanations are consistent
with the data.

First, our finding may reflect a genuine editorial choice by
English-language outlets to frame the outbreak as a structural
policy failure rather than a partisan issue---a framing consistent
with the recent domestic literature~\cite{hossain2026}. Second, the
systemic causes were, by early 2026, empirically visible in the
epidemiological record: a missed quadrennial campaign is a
programme-level failure that spans administrations and resists
personalization to any single actor. Third, and more tentatively,
English-language media in Bangladesh operated during this period
under documented pressures on press
freedom~\cite{bbc2024mediaattacks}; systematic individual-blame
coverage of any government carries reputational and, at times,
operational risk. We stress that our headline-level data cannot
adjudicate among these explanations, and we do not claim to
demonstrate causal editorial self-restraint. What our data do show
is that even under conditions where personalized political blame
would be plausible, English-language coverage overwhelmingly
located fault in the health system rather than in named actors---a
pattern worth further investigation with newsroom-level and
article-body evidence.

\subsection{Methodological Contribution}

The pipeline demonstrates a transferable approach for outbreak
media analysis in data-constrained settings: reproducible
Wayback-anchored retrieval, a locked codebook, and validated LLM
annotation. Strong stance agreement with humans
($\kappa{=}0.89$) is consistent with growing evidence that LLMs
can serve as reliable annotators for CSS when validated against a
gold standard~\cite{gilardi2023,heseltine2024}. The
headline-match retrieval adaptation broadens applicability to
outlets with opaque URLs without altering search scope.

\subsection{Limitations}

The corpus is restricted to seven national English-language
outlets; Bangla coverage is excluded, and three candidate outlets
were dropped for technical reasons. Blame, though analytically
central, is infrequent (37 headlines), yielding small phase-level
cells that preclude reliable significance testing of its
trajectory. Headline-level analysis captures salient
agenda-setting framing but not article-body nuance. The study
covers a single outbreak in a single country, and June coverage
was partially archived. Sentiment was collapsed to binary after
human coders diverged on the three-way scheme; this is a
defensible simplification but sacrifices fine gradation. Finally,
we report reliability against a two-coder human-adjudicated gold
standard; systematic comparison across multiple LLM providers and
prompt conditions is left to future work.

\section{Conclusion}

Applying a validated LLM-annotation pipeline to a
reproducibility-first corpus of 403 headlines across seven
outlets, English-language Bangladeshi coverage of the 2026
measles outbreak became significantly more negative and more
threat-focused as cumulative mortality rose, briefly foregrounded
institutional reassurance during the vaccination campaign, and
attributed blame sparingly and systemically. Media framing
tracked the outbreak's cumulative cost more closely than its
real-time epidemiological course, and---despite a politically
charged context---amplified threat far more than it assigned
political fault. The method offers a transparent, transferable
approach to characterizing media responses to health emergencies
in the LMIC settings where such evidence is most needed.

%---------------------------------------------------------------
% References
%---------------------------------------------------------------


\begin{thebibliography}{99}

\bibitem{downs1972}
Downs, A.: Up and down with ecology: the ``issue-attention cycle.''
The Public Interest \textbf{28}, 38--50 (1972)

\bibitem{shih2008}
Shih, T., Wijaya, R., Brossard, D.: Media coverage of public health
epidemics: linking framing and issue attention cycle toward an
integrated theory of print news coverage of epidemics. Mass
Communication and Society \textbf{11}(2), 141--160 (2008)

\bibitem{ogbodo2020}
Ogbodo, J.N., et al.: Communicating health crisis: a content
analysis of global media framing of COVID-19. Health Promotion
Perspectives \textbf{10}(3), 257--269 (2020)

\bibitem{schafer2019}
Sch\"afer, M.S.: The notorious GRP: investigating the framing of
Austria's largest measles outbreak. Journalism Studies
\textbf{20}(13), 1896--1916 (2019)

\bibitem{gilardi2023}
Gilardi, F., Alizadeh, M., Kubli, M.: ChatGPT outperforms
crowd-workers for text-annotation tasks. Proceedings of the
National Academy of Sciences \textbf{120}(30), e2305016120 (2023)

\bibitem{heseltine2024}
Heseltine, M., Clemm von Hohenberg, B.: Large language models as
a substitute for human experts in annotating political text.
Research \& Politics \textbf{11}(1) (2024).
\doi{10.1177/20531680241236239}

\bibitem{walker2025}
Walker, J.T., Angst, C.M., Sanchez, T.W.: When large language
models fail: stance detection in policy contexts. PLOS ONE
\textbf{20}(2), e0318234 (2025)

\bibitem{dghs2026}
Directorate General of Health Services (DGHS), Bangladesh:
Measles situation report, 26 June 2026. Government of the People's
Republic of Bangladesh, Dhaka (2026).
\url{https://old.dghs.gov.bd/}

\bibitem{who2026}
World Health Organization: Disease outbreak news: measles ---
Bangladesh, 23 April 2026 (2026).
\url{https://www.who.int/emergencies/disease-outbreak-news}

\bibitem{idea2024bangladesh}
International IDEA: Bangladesh --- August 2024. Democracy Tracker
(2024).
\url{https://www.idea.int/democracytracker/report/bangladesh/august-2024}

\bibitem{bmj2026bangladesh}
Rahman, M., et al.: Political turmoil, disrupted immunisation
services, and the resurgence of measles in Bangladesh. BMJ
\textbf{2026}, s819 (2026). \doi{10.1136/bmj.s819}

\bibitem{prothomalo2026shortage}
Prothom Alo: Nationwide outcry for vaccines: shortages persist
despite payment. Prothom Alo English Edition (2026).
\url{https://en.prothomalo.com/bangladesh/yqni6rdl84}

\bibitem{plansrubio2025}
Plans-Rubi\'o, P.: Measles vaccination coverage and anti-measles
herd immunity levels in the World and WHO Regions worsened from
2019 to 2023. Vaccines \textbf{13}(2), 157 (2025).
\doi{10.3390/vaccines13020157}

\bibitem{hossain2026}
Hossain, S., Ahmed, A., Hossain, M.S.: Resurgence of measles in
Bangladesh amid a global upsurge: an urgent call for emergency
public health response. Tropical Medicine and Health
\textbf{54}, 96 (2026). \doi{10.1186/s41182-026-00982-y}

\bibitem{who2026don598}
World Health Organization: Measles --- Bangladesh, 2026. Disease
Outbreak News (2026).
\url{https://www.who.int/emergencies/disease-outbreak-news/item/2026-DON598}

\bibitem{iyengar1991}
Iyengar, S.: Is Anyone Responsible? How Television Frames Political
Issues. University of Chicago Press, Chicago (1991)

\bibitem{bbc2024mediaattacks}
BBC News: Bangladesh journalists face attacks and legal pressure
amid political transition (2024).
\url{https://www.bbc.com/news/articles/c74xge7z8gqo}

\end{thebibliography}
\end{document}